\catcode`\@11
\documentclass[11pt]{article}
\usepackage{times}\usepackage{cite} \usepackage{amssymb}
\usepackage{pifont}\usepackage{amsmath}\usepackage{amscd}
\usepackage{eepic}\usepackage{psfig}\usepackage{euscript}
\numberwithin{equation}{section}
\newcommand\remove[1]{}
\newcommand\preprint[1]{\vspace{-1in}\vtop{\null\hfill
\parbox[t]{1.6in}{\small\sc #1\\\null}}
\vskip .5in\bigskip\normalfont}

\renewcommand\bar\overline
\newcommand\cf[4]{\bibitem{#1}{#2}.~{\it #3};~{#4}.}
\newcommand\cff[5]{{\bibitem{#1}{#2}.~{\it #3};~{#4};~{#5}}.}

\def\ch#1{\qopname\relax o{\sf ch}{[#1]}}
\def\chun#1{\qopname\relax o{\sf ch}{\!{}_1[#1]}}
\def\chdu#1{\qopname\relax o{\sf ch}{\!{}_2[#1]}}
\def\cstar{{\mathbb{C}^{\star}}}

\def\C{{\mathbb{C}}}
\def\cp#1{{\mathbb{P}}^{#1}}
\def\D{{\goth{D}}}
\def\diag#1{\qopname\relax o{\sf diag}{(#1)}}

\def\dual#1{\overset \vee {#1}}
\newcommand\eg{{\slshape e.g.~}}
\def\eq#1{(\ref{#1})}

\def\Gm#1{\Gamma\left(#1\right)}
\def\goth#1{{\mathfrak #1}}
\renewcommand\hat{\widehat}
\def\hom#1{\qopname\relax o{\EuScript{H}om({#1})}}
\def\id#1{{\mathbb{I}}_{#1}}
\newcommand\ie{{\slshape i.e.~}}

\def\ker#1{\qopname\relax o{Ker}~{#1}}

\def\map{\longmapsto}
\def\m{{\mathfrak m}}
\def\M{{\cal M}}
\def\bm{\mathbf{M}}
\def\bn{\mathbf{N}}

\def\pa{\partial}
\def\pic#1{\operatorname{\EuScript{P}{ic}}({#1})}

\def\rt{\longrightarrow}
\def\R{{\mathbb{R}}}

\def\t#1{\Theta_{#1}}
\def\td#1{\qopname\relax o{\sf td}{[#1]}}
\def\tilde{\widetilde}

\def\viz{{\slshape viz.~}}
\def\wcp#1#2{{\mathbf{P}}(\C^{#1})^{[#2]}}
\def\Z{{\mathbb{Z}}}

\def\vt{\vartheta}
\newcommand\piorb[2]{\gdef\ciclo{#1}%
\ifx\@empty\ciclo{\overset \infty \Pi_{#2}}\else
{\overset \infty  \Pi_{#2}^{[#1]}}\fi}
\newcommand\pil[2]{\gdef\lpciclo{#1}%
\ifx\@empty\lpciclo{\Pi_{#2}}\else
{\Pi_{#2}^{[#1]}}\fi}
\newcommand\vpiorb[2]{\gdef\vciclo{#1}%
\ifx\@empty\vciclo{\overset \infty \varpi_{#2}}\else
{\overset \infty \varpi_{#2}^{[#1]}}\fi}
\newcommand\vpil[2]{\gdef\lciclo{#1}%
\ifx\@empty\lciclo{\varpi_{#2}}\else
{\varpi_{#2}^{[#1]}}\fi}
\DeclareFontFamily{U}{rsf}{}
\DeclareFontShape{U}{rsf}{m}{n}{
  <5> <6> rsfs5 <7> <8> <9> rsfs7 <10-> rsfs10}{}
\DeclareMathAlphabet\Scr{U}{rsf}{m}{n}
\def\sheaf#1{{\Scr #1}}
\newcommand\atmp[3]{Adv. Theor. Math. Phys. {\bf #1}~(#2)~#3}
\newcommand\alg[1]{alg-geom/{#1}}
\newcommand\cmp[3]{Comm. Math. Phys.{\bf #1}~({#2})~{#3}}

\newcommand\fnaa[3]{Func. Anal. and Appl. {\bf {#1}}~({#2})~{#3}}
\newcommand\hepth[1]{hep-th/{#1}}

\newcommand\jams[3]{J. Amer. Math. Soc. {\bf {#1}}~({#2})~{#3}}

\newcommand\jhep[3]{JHEP {\bf {#1}}~({#2})~{#3}}
\newcommand\ag[1]{{ math.AG/{#1}}}
\newcommand\npb[3]{Nucl.~Phys. {\bf B{#1}}~({#2})~{#3}}

\newcommand\plb[3]{Phys. Lett. {\bf B{#1}}~({#2})~{#3}}

\begin{document}
\title{\preprint{USITP-01-04\\hep-th/0102146}
Fractional Branes on a Non-compact Orbifold}
\author{}
\date{
Subir Mukhopadhyay
\thanks{E--mail:~ subir@physto.se} 
\\{\small{\sl 
Institute of Theoretical Physics, University of Stockholm,\\Box 6730, S-113 85, 
Stockholm, Sweden}
}
\\{\sl\&}\\ Koushik Ray
\thanks{E--mail:~koushik@ictp.trieste.it}
\\{\small
{\sl  the Abdus Salam International Center for Theoretical Physics\\
Strada Costiera, 11 -- 34014~Trieste, Italy}
}}%
\maketitle
\thispagestyle{empty}
\vfil
\begin{abstract}
\noindent Fractional  branes on  the non-compact  orbifold $\C^3/\Z_5$ are 
studied.  First,  the boundary  state  description of  the fractional branes 
are obtained.  The open-string  Witten index calculated  using these 
states  reproduces  the  adjacency  matrix  of the  quiver of $\Z_5$.  Then, 
using the toric crepant resolution of the orbifold $\C^3/\Z_5$ and invoking 
the local mirror principle, B-type branes wrapped on the holomorphic cycles 
of the resolution are  studied. The  boundary 
states corresponding to the five fractional branes are  identified as bound 
states of BPS D-branes  wrapping the 0-, 2- and 4-cycles in the exceptional  
divisor of the resolution of $\C^3/\Z_5$. 
\end{abstract}
\clearpage
\clearpage
\section{Introduction}\label{intro}
D-branes wrapping supersymmetric cycles embedded in a variety purvey simple 
examples of curved D-branes.  For special values of the moduli of string 
theory, as a cycle shrinks to null volume, leading, perhaps, to an orbifold 
singularity, 
the branes wrapped on it may give rise to states which are pinned to a 
point in the transverse space, perambulating only the Coulomb branch.
Their charges and masses are certain fractions of that of a BPS D-brane; 
hence the neologism, \emph{fractional brane}\cite{doug,dgomis}.
 Fractional branes on orbifolds  
appear also in the guise of boundary states of a perturbative superconformal 
field theory, graded by the irreducible representations of the quotienting 
group. The two descriptions are nonetheless 
related and pertain to different points in the
moduli space of the conformal field theory (CFT). Confining our discussion to
orbifolds $\C^3/G$ for some discrete subgroup $G$ of $SL(3,\C)$, the wrapped
branes furnish a valid description of these states at a point where the
exceptional surfaces of the resolution of the orbifold are of large volume,
while the CFT describes the states at the point where the exceptional
surfaces shrink to vanishing volume giving rise to the orbifold singularity.
We shall refer to the latter point as the \emph{orbifold point} in the sequel.
Discerning the guises of D-branes in different regions and exploring their
interrelations are of import in obtaining geometric interpretations of the
conformal field theoretic states, as well as clarifying several aspects of
gauge field theories. Such studies have also
provided physical realisations of the mathematical connection between
representations of discrete groups and the cohomology of resolved orbifolds,
namely, the McKay correspondence, in addition to providing evidence for mirror
symmetry and its generalisations.

Different aspects of the identification of wrapped branes at the orbifold limit
and the boundary states have been studied recently, from various
points of view 
\cite{dg,mohri,dfr1,dfr2,dd,bdlr,kllw,sch,jay1,jay2,jay3,tom,mayr,doug0}, 
confirming the geometric provenance of the fractional branes.
Examples of threefolds on which the above scenario has been tested
include D0-branes on blown up $\C^3/G$, for $G=\Z_3,\Z_4,\Z_6$. 
The fractional D0-branes corresponding to $G=\Z_N$ have 
$1/N^{\mathrm{th}}$ the mass and charge of a D0-brane. 
In the large volume limit the fractional branes 
manifest themselves as 
bound states of branes wrapped on the holomorphic cycles of the
resolution. For example, in the case 
of $\C^3/\Z_3$\cite{dg, dfr1,mohri}, there are three fractional brane states,
carrying one-third the mass and charge of a D0-brane. 
The exceptional surface of the resolution is a $\cp{2}$.
In the large volume limit, BPS branes wrap on the holomorphic 
cycles of the $\cp{2}$. Thus,
the fractional branes owe their origin to wrapped BPS D4-, D2- and
D0-brane bound states. More precisely, they are
identified with the trivial line bundle $\sheaf{O}$, 
the tautological line bundle $\sheaf{O}(-1)$ (up to signs) 
and an exceptional bundle of rank two on $\cp{2}$.
The analysis has been generalised to other cases, especially to the partial 
resolutions of $\C^3/\Z_4$ and $\C^3/\Z_6$\cite{mohri}. 
These necessitated considering D4-branes wrapped on weighted projective spaces,
$\wcp{2}{112}$ and $\wcp{2}{123}$ and hence bundles on these. The 
map between the fractional branes and the bundles on the divisors 
of the partial resolutions has been obtained in both these cases. 
Similar analyses on compact Calabi-Yau manifolds\cite{bdlr,kllw,sch}
yield a large volume D-brane
interpretation of Gepner model states \cite{rs,gut,AS,abpss}.

A feature common to the examples of threefolds mentioned above is that the
Newton polygon corresponding to each of them 
is reflexive, \ie contains a single point in its interior. 
In this article we extend certain aspects of such studies to another
non-compact threefold, namely the crepant resolution $\mathbf{X}$
of $\C^3/G$, for $G=\Z_5\subset SL(3,\C)$.
\begin{eqnarray}
\begin{CD}
&&&\mathbf{X}\\
&&&@VVV&&\\
{\C^3} @>{G}>>&{\C^3/G}&&
\end{CD}
\nonumber
\end{eqnarray}
The Newton polygon corresponding to the threefold $\mathbf{X}$ 
has two points in its interior. Thus, our result will furnish further 
evidence for the afore-mentioned scenario in a more general situation.

We shall consider D0-branes in Type-IIA theory reduced on
$\C^3/G$.  We start by performing the quotienting by $G=\Z_5$ on 
the partition function of
the ${\cal N}=2$ theory. This allows us to identify the boundary states
corresponding to the fractional branes.
The identification of wrapped branes corresponding to these states in the
resolution $\mathbf{X}$ closely parallels the example of $\C^3/\Z_3$ \cite{dg}. 
The five fractional branes on the orbifold form an orbit of the 
quantum $\Z_5$ symmetry. Since only the even-dimensional cohomologies of the
resolution are non-trivial, the fractional D0-branes may occur only by 
wrapping of B-type branes in the Type-IIA theory on the holomorphic cycles 
in the large volume limit. 
We shall identify the set of wrapped branes which form an identical
orbit of the monodromy around the orbifold point.
In order to find the orbit in the large volume limit we shall obtain the 
volume of the cycles over the K\"ahler moduli space.
This require determining the 0-, 2- and 4-cycle periods of $\mathbf{X}$, of
which the 2- and 4-cycles have to be Legendre dual of each other with respect
to the prepotential. The duals can not be identified correctly without the
knowledge of the triple-intersection numbers of the divisors of $\mathbf{X}$.
The triple-intersection numbers as calculated on $\mathbf{X}$ are known to be
receive world-sheet instanton corrections. 
Considerations of world-sheet instanton corrections are avoided 
by taking recourse to the mirror geometry and hence working with
the complex structure deformations of the mirror. For the non-compact 
threefold $\mathbf{X}$ at hand, the relevant version of mirror symmetry is the
local one. We obtain the local mirror which is a
Riemann surface of genus two for the present example.
The large volume limit of the K\"ahler moduli space corresponds to the large
complex structure limit ({\sc lcsl}) in the mirror, while the orbifold limit is
given by the vanishing of these deformations. 

We then study the semi-periods \cite{adj} given by the 
${\cal A}$-hypergeometric system of solutions to the 
associated GKZ system\cite{gkz,ode,hly,hosono,horja} 
in the {\sc lcsl}, which is identified as the point of maximal
degeneracy of the indicial equations of the GKZ system.
The solutions are then
continued analytically to the orbifold point and we find out the monodromy
of periods around that point. There are five charged states in this regime
which form periods of the monodromy matrix. By virtue of being of equal mass,
these can be identified with the fractional branes. The BPS-brane content of
these states can be read off from the charges of 0-, 2- and 4-cycle periods.

The plan of the paper is as follows. First, we  obtain the boundary 
states for the fractional branes on the orbifold in \S\ref{opnstr}. We then 
describe the toric resolution of the orbifold, its local
mirror geometry and describe the GKZ system for the mirror in \S\ref{gkzsec}.
The periods in the {\sc lcsl}  are obtained by solving the GKZ system 
in \S\ref{lcslsec}.
Finally, in \S\ref{orbsec}, we analytically continue the periods to 
the orbifold 
point and find out the monodromies around the orbifold point which is 
then used for the identification of the 
fractional branes with the wrapped branes at large volume. 
We summarise the results in \S\ref{concl}.
\section{Open strings on the orbifold $\C^3/G$}
\label{opnstr}
The orbifold $\C^3/\Z_5$ is defined by the action of $\Z_5$ on $\C^3$ through
the defining representation $\diag{\omega,\omega,\omega^3}$ on the three
complex coordinates, 
\begin{equation} 
\Z_5: (z_1, z_2, z_3)\map (\omega^{a_1} z_1, \omega^{a_2} z_2, 
\omega^{a_3} z_3),
\quad (z_1, z_2, z_3)\in \C^3,
\end{equation}
where $\omega$ is a fifth root of unity:
$\omega = e^{\frac{2\pi i}{5}}$, $a_1=a_2=1$ and $a_3=3$.
The orbifold is also denoted $\frac{1}{5}{[113]}$. 
In this section we shall first consider the string partition function in the 
open string channel.
The action of the group $G$ is effected on the partition function.
Then using the modular properties of theta functions, the same is
translated to expressions in terms of closed string variables and the partition 
function is
expressed as a transition amplitude between boundary states. Finally, we
calculate the open string Witten index as the transition amplitude 
between the $I^{\mathrm{th}}$ and the $J^{\mathrm{th}}$ states 
in the R-sector. Collecting them in a
matrix produce the adjacency matrix of the Quiver diagram of $G$.
\\{\sl {\bf NB:}~Throughout this section
a lone $\prod$ abbreviates $\displaystyle\prod_{n=1}^{\infty}$.} \\
\subsection{Partition function in the open string channel}
D-branes carry the Chan-Paton charges of open strings. Thus, in addition to
specifying the action of $G$ on $\C^3$, we have to specify its action on the
Chan-Paton degrees of freedom. The action of $G$ on the Chan-Paton
indices, leading to one single D0-brane \cite{dm,dgm} is consistent within 
string theory only if  $G$ acts on the Chan-Paton matrices through its 
regular representation, namely ${\cal R}=\bigoplus_Id_IR_I$, where $R_I$
denotes the $I^{\mathrm{th}}$ irreducible representation, $I=1,{\ldots},|G|$ and
$|G|$ denotes the order of the group. Thus, for the case at hand, $|G|=5$.
The multiplicities $d_I=1$ for all $I$. However, we shall keep the $d_I$
for book-keeping; the formulas later in this section are actually 
valid in more general situations.
Furthermore, since $G$ is cyclic, the elements of $G$ can be written as $g^m$,
$m=0,{\ldots},|G|-1$, for an element $g$ of $G$. 

We consider the open string connecting $I^{\mathrm{th}}$ and 
$J^{\mathrm{th}}$ representations.
The $G$-projected partition function in a sector
$\goth{s}$ can be written as 
\begin{equation}
{\cal Z}_{\goth{s}} =
\frac{d_Id_J}{|G|}\sum_{m=0}^{|G|-1}\bar{\chi}^I(m)\chi^J(m)
\mathrm{Tr}_{\goth{s}}\left(g^m e^{-2{\tau}H_o}\right),
\end{equation}
where $\goth s$ is either the NS or the R sector, in which the trace is taken
and $\chi^I(m)$ denotes the $I^{\mathrm{th}}$ character of $g^m$. Moreover,
$H_o$ denotes the open string Hamiltonian 
and is given by
\begin{equation}
H_o = \pi p^2 + \pi\sum_{\mu=0,{\ldots},7} (\sum_{n=1}^\infty 
\alpha_{-n}^\mu\alpha_n^\mu + \sum_{r>0} r\psi_{-r}^\mu \psi_r^\mu) + \pi c_0, 
\end{equation} 
where $\alpha$ refers to the oscillators, $p$ denotes the momentum
and $c_0$ is a constant\cite{dg}.

We now implement the action of $G$ on the different sectors of the partition
function. The partition function in the NS-sector is 
\begin{equation}
\begin{split}
{\cal{Z}}_{\scriptscriptstyle\mathrm{NS}}(0,\tau)&=\frac{1}{
\eta(\tau)^8}
\left[\frac{\vt_3(\tau)}{\eta{(\tau)}}\right]^4 \\
&= q^{-1/2} \frac{\prod (1+q^{n-1/2})^8}{\prod (1-q^n)^8},
\end{split}
\end{equation}
where $q= e^{2i\pi\tau}$.
Reduced on $\C^3$ and projected by ${g}^m$, it takes the form
\begin{equation}
\begin{split}
{\cal Z}_{\scriptscriptstyle\mathrm{NS}}(m,\tau) &= 
q^{-1/2}\left[\frac{\prod(1+q^{n-1/2})}
{\prod(1-q^n)}\right]^2
\prod_{i=1}^3\frac{\prod(1+\omega^{a_i}q^{n-1/2})
(1+\bar{\omega}^{a_i}q^{n-1/2})}
{\prod(1-\omega^{a_i}q^n)(1-\bar{\omega}^{a_i}q^n)},\\
&= 8\thinspace\frac{\vt_3(\tau)}{\eta(\tau)^3}\prod_{i=1}^3
\frac{\vt_3(\frac{ma_i}{N}|\tau)}{\vt_1(\frac{ma_i}{N}|\tau)}
\thinspace\sin\left(\frac{\pi m a_i}{N}\right).
\end{split}\end{equation}
Similarly, from the unprojected partition function of the 
$(-1)^F\mathrm{NS}$-sector, \viz
\begin{equation}
{\cal Z}_{(-1)^F\scriptscriptstyle\mathrm{NS}}(0) = 
\frac{1}{\eta(\tau)^8}\thinspace
\left[\frac{\vt_4(\tau)}{\eta(\tau)}\right]^4,
\end{equation}
we derive the following after projection by ${g}^m$
\begin{equation}
{\cal Z}_{(-1)^F\scriptscriptstyle\mathrm{NS}}(m,\tau) = 
8\thinspace\frac{\vt_4(\tau)}{\eta(\tau)^3}\prod_{i=1}^3
\frac{\vt_4(\frac{ma_i}{N}|\tau)}{\vt_1(\frac{ma_i}{N}|\tau)}
\thinspace\sin\left(\frac{\pi m a_i}{N}\right).
\end{equation}
The partition function in the R-sector, on the other hand, 
after projection, takes the form
\begin{equation}
{\cal Z}_{\scriptscriptstyle\mathrm{R}}(m,\tau) = 
8\thinspace\frac{\vt_2(\tau)}{\eta(\tau)^3}\prod_{i=1}^3
\frac{\vt_2(\frac{ma_i}{N}|\tau)}{\vt_1(\frac{ma_i}{N}|\tau)}
\thinspace\sin\left(\frac{\pi m a_i}{N}\right).
\end{equation}
Collecting contributions from the different sectors after projection, 
the partition function on $\C^3/G$ reads
\begin{equation}
\tilde{{\cal Z}}(m,\tau) = 
{\cal Z}_{\scriptscriptstyle\mathrm{NS}}(m,\tau)-  
{\cal Z}_{(-1)^F\scriptscriptstyle\mathrm{NS}}(m,\tau)  -
{\cal Z}_{\scriptscriptstyle\mathrm{R}}(m,\tau).
\end{equation}
The total partition function is obtained by summing this 
over all the elements of the group,
\begin{equation}
\label{ztot}
{\cal Z} = \frac{d_Id_J}{|G|}\sum_{m=0}^{N-1}\hat{\cal Z}(m,\tau)
\thinspace\bar{\chi}^I(m)\chi^J(m),
\end{equation}
where 
\begin{equation}
\hat{\cal Z}(m,\tau)=\frac{V}{2\pi}\int\frac{dt}{4\sqrt{2}t^{3/2}}
\tilde{\cal Z}(m,\tau) 
\end{equation}
and $t=-i\tau$.
\subsection{Boundary states}
Boundary states provide a closed string theoretic description of the CFT and 
can be obtained from the partition function in the open string sector
\cite{dg,roose,taka}. The 
basic idea is to find the Ishibashi states in the closed string CFT, which 
satisfy certain linearized boundary conditions. 
However all such states do not correspond to a state describing a D-brane. 
In order to qualify as a physical state, they have to satisfy the Cardy's
conditions that the closed string scattering amplitude should 
factorise to appropriate sectors in the open string channel. Moreover,
the physical states will have to survive the GSO projection.
In the case of orbifold, due to conformal invariance we have an additional 
factorization condition on the boundary states \cite{dg}. 

In order to identify the boundary state, let us write the open string 
partition function \eq{ztot} in terms of closed string variables
as 
\begin{equation}\label{ztotcl}
{\cal Z} = \frac{d_Id_J}{|G|}\sum_{m=0}^{N-1}
\hat{\cal Z}_{\mathrm{cl}}(m,l)
\thinspace\bar{\chi}^I(m)\chi^J(m),
\end{equation}
where 
\begin{equation}
\hat{\cal Z}_{\mathrm{cl}}(m,l)
=\frac{iV}{2\pi}\int\frac{dl}{l^{3/2}}
\tilde{\cal Z}_{\mathrm{cl}}(m,l), 
\end{equation}
and         
\begin{equation}
\begin{split}
\tilde{\cal Z}_{\mathrm{cl}}(m,l) &= 
\thinspace\frac{\vt_3(2il)}{\eta(2il)^3}\prod_{i=1}^3
\frac{\vt_3(\frac{-2ma_il}{N}|2il)}{\vt_1(\frac{-2ma_il}{N}|2il)}
\thinspace\sin\left(\frac{\pi m a_i}{N}\right)\\
&\quad -\thinspace\frac{\vt_4(2il)}{\eta(2il)^3}\prod_{i=1}^3
\frac{\vt_4(\frac{-2ma_il}{N}|2il)}{\vt_1(\frac{-2ma_il}{N}|2il)}
\thinspace\sin\left(\frac{\pi m a_i}{N}\right)
\\&\quad -\thinspace\frac{\vt_2(2il)}{\eta(2il)^3}\prod_{i=1}^3
\frac{\vt_2(\frac{-2ma_il}{N}|2il)}{\vt_1(\frac{-2ma_il}{N}|2il)}
\thinspace\sin\left(\frac{\pi m a_i}{N}\right),
\end{split}
\end{equation}
using modular transformation properties of the theta functions \cite{dg}.

In the closed string channel we have the Ishibashi states satisfying 
appropriate Neumann-Dirichlet conditions with the different spin structures.
They are the coherent states obtained by acting exponentiated sum of 
oscillators on the vacuum. For every twisted sector there is an 
Ishibashi state with a given 
spin structure from NS-NS and R-R sectors.
Boundary states are obtained by writing \eq{ztotcl} as transition amplitudes
between such states. This is achieved using the expressions of the transition
amplitudes between Ishibashi states in terms of theta functions in each 
twisted sector in the closed string channel \cite{dg},
\begin{equation}
\begin{split}
\int dl 
\langle +,m|\thinspace e^{-l{{H}}_{\mathrm{cl}}}|+,m
\rangle_{\scriptscriptstyle\mathrm{NS-NS}} &=
\int dl 
\langle -,m|\thinspace e^{-l{H}_{\mathrm{cl}}}|-,m
\rangle_{\scriptscriptstyle\mathrm{NS-NS}} \\
&= i{\sf N}_m^2\int\frac{dl}{l^{3/2}}\thinspace\vt_3(2il)
\prod_{i=1}^{3}
\frac{\vt_3(\frac{-2ma_il}{N}|2il)}{\vt_1(\frac{-2ma_il}{N}|2il)},
\end{split}
\end{equation}
\begin{equation}
\begin{split}
\int dl 
\langle +,m|\thinspace e^{-l{H}_{\mathrm{cl}}}|-,m
\rangle_{\scriptscriptstyle\mathrm{NS-NS}} &=
\int dl 
\langle -,m|\thinspace e^{-l{H}_{\mathrm{cl}}}|+,m
\rangle_{\scriptscriptstyle\mathrm{NS-NS}} \\
&= i{\sf N}_m^2\int\frac{dl}{l^{3/2}}\thinspace\vt_2(2il)
\prod_{i=1}^{3}
\frac{\vt_2(\frac{-2ma_il}{N}|2il)}{\vt_1(\frac{-2ma_il}{N}|2il)},
\end{split}
\end{equation}
\begin{equation}
\begin{split}
\int dl 
\langle +,m|\thinspace e^{-l{H}_{\mathrm{cl}}}|+,m
\rangle_{\scriptscriptstyle\mathrm{RR}} &=
\int dl 
\langle -,m|\thinspace e^{-l{H}_{\mathrm{cl}}}|-,m
\rangle_{\scriptscriptstyle\mathrm{RR}} \\
&= -i{\sf N}_m^2\int\frac{dl}{l^{3/2}}\thinspace\vt_4(2il)
\prod_{i=1}^{3}
\frac{\vt_4(\frac{-2ma_il}{N}|2il)}{\vt_1(\frac{-2ma_il}{N}|2il)},
\end{split}
\end{equation}
with the normalisation determined by demanding equality of the expressions in
the two channels as 
\begin{equation}
{\sf N}_m^2 = 
\frac{d_I^2}{|G|}
\prod_{i=1}^3\sin\left(\frac{\pi ma_i}{N}\right)|\chi^I(m)|^2,
\end{equation}
from which we can find out the normalizations of the Ishibashi states with 
different spin structures associated to a representation $I$.
In these formulas ${H}_{\mathrm{cl}}$ denoting the closed string Hamiltonian.

Boundary states are then obtained by comparing the different terms in the 
partition function \eq{ztotcl} and the above amplitudes, both now in the 
closed string variables.
\begin{equation}
|{\cal B},m,I\rangle = \frac{1}{2}\left( 
|+,m,I\rangle_{\scriptscriptstyle\mathrm{NS-NS}}
-|-,m,I\rangle_{\scriptscriptstyle\mathrm{NS-NS}}
+|+,m, I\rangle_{\scriptscriptstyle\mathrm{R-R}}
+|-,m, I\rangle_{\scriptscriptstyle\mathrm{R-R}}\right),
\end{equation}
for $m=0,{\ldots},N-1$.
The index $m$ labels the state that pertains to the
element  $g^m$ of $G$  and $I$ the irreducible
representation. The Chan-Paton indices are collected in $\cal B$.
The signs ($\pm$) stand for the respective spin structures.
We now write the states in 
a basis corresponding to the nodes of the quiver diagram, 
which, in turn, corresponds to the irreducible 
representations as \cite{taka}
\begin{equation}
\label{bdst}
\mid {\cal B},I\rangle =\sum_{m=0}^{|G|-1}\mid {\cal B},m\rangle .
\end{equation}
Let us point out that the boundary states \eq{bdst} are graded by the 
characters of the representation of $G$ through $I$. This one-to-one
correspondence between the
boundary states and the characters of representation is the crucial ingredient
in the identification of these boundary states as branes wrapping homology
cycles in the resolution of the orbifold, which may be thought of as a
physical realisation of the McKay correspondence \cite{reid}.
\subsection{The open-string Witten Index}
\label{witten}
Let us now consider the partition function in the RR-sector in a little more
detail. Extracting the contribution from the RR-sector in 
$\hat{\cal Z_{\scriptscriptstyle\mathrm{cl}}}$, we have
\begin{equation}
{\cal Z}_{\scriptscriptstyle\mathrm{R}}(\tau) = 
\frac{8d_Id_J}{|G|}\thinspace\frac{\vt_2(\tau)}{\eta(\tau)^3}
\sum_{m=0}^{N-1}
\prod_{i=1}^3\frac{\vt_2(\frac{ma_i}{N}|\tau)}{\vt_1(\frac{ma_i}{N}|\tau)}
\thinspace\sin\left(\frac{\pi m a_i}{N}\right).
\end{equation}
Accordingly, the contribution to $\cal Z$ of \eq{ztotcl} from the  
massless part of ${\cal Z}_{\scriptscriptstyle\mathrm{R}}(\tau)$, 
after GSO projection, written as a matrix in the $I$ indices, 
yields the matrix of Witten indices \cite{roose,taka},
\begin{equation}
{\cal W}_{IJ} = 
\frac{8d_Id_J}{|G|}\sum_{m=0}^{N-1}\prod_{i=1}^3 (-1)^m 
\cos\left(\frac{\pi m a_i}{N}\right)\bar{\chi}^I(m)\chi^J(m).
\end{equation}
Using $d_I=1$ for all $I$ for the regular representation and the
character table for $G=\Z_5$, shown in Table~1, we obtain the matrix $\cal W$
for the present example.

\vspace{7mm}
\hbox{%
\parbox[h]{.5\textwidth}{
\begin{center}
\begin{tabular}{|l|ccccc|}
\hline
$\substack{\;\;\;\;m\\I\;\;\;\;}$
&$\substack{0\\{} }$ &$\substack{1\\{} }$ &$\substack{2\\{} }$
&$\substack{3\\{} }$ &$\substack{4\\{} }$\\\hline
0&1&1&1&1&1\\
1&1&$\omega$ & $\omega^2$ & $\omega^3 $& $\omega^4$\\
2&1&$\omega^2$ & $\omega^4$ & $\omega$   & $\omega^3$\\
3&1&$\omega^3$ & $\omega  $ & $\omega^4 $& $\omega^2$\\
4&1&$\omega^4$ & $\omega^3$ & $\omega^2$ & $\omega $\\
\hline 
\end{tabular}
\end{center}
}\parbox[h]{.5\textwidth}{%
\setlength{\unitlength}{2647sp}%
\begin{center}
\begin{picture}(2752,2750)(4625,-6536)
\thinlines
\put(6001,-3961){\circle*{336}}\put(5401,-6361){\circle*{336}}
\put(6601,-6361){\circle*{336}}\put(7201,-5161){\circle*{336}}
\put(4801,-5161){\circle*{336}}
\put(5551,-6286){\vector( 1, 0){900}}
\put(5551,-6436){\vector( 1, 0){900}}
\put(6751,-6286){\vector( 1, 2){480}}
\put(6611,-6215){\vector( 1, 2){480}}
\put(7231,-4998){\vector(-1, 1){1050}}
\put(7110,-5116){\vector(-1, 1){1020}}
\put(5851,-3961){\vector(-1,-1){1050}}
\put(5926,-4111){\vector(-1,-1){975}}
\put(4921,-5251){\vector( 1,-2){480}}
\put(4776,-5324){\vector( 1,-2){480}}
\put(4951,-5161){\vector( 1, 0){2100}}
\put(5476,-6211){\vector( 1, 4){525}}
\put(6001,-4111){\vector( 1,-4){525}}
\put(7051,-5236){\vector(-3,-2){1520}}
\put(6451,-6286){\vector(-4, 3){1500}}
\end{picture}
\end{center}
}
} 
\hbox{%
\parbox[h]{.5\textwidth}{%
\centerline{Table:~{\thetable}.
~~\small\sl Characters $(\chi^I(m))$ of $G=\Z_5$}
}
\parbox[h]{.5\textwidth}{%
\centerline{Figure~{\thefigure}.~~\small\sl Quiver diagram for $G=\Z_5$}
}
}\vspace{7mm} %
The matrix $\cal W$ reproduces the adjacency matrix of the quiver diagram 
of $G=\Z_5$, \viz
\begin{equation}
\label{quivermat}
{\cal C}=\begin{pmatrix}
0 & 2 & -1 & 1 &-2\\
-2&0&2&-1&1\\
1&-2&0&2&-1\\
-1&1&-2&0&2\\
2&-1&1&-2&0
\end{pmatrix},
\end{equation}
up to signs. Thus, ${\cal W}=|\cal C|$. 
Let us emphasise that the boundary states that yield the adjacency matrix are
on the \emph{orbifold} $\C^3/\Z_5$. Later, we shall find that the same
adjacency matrix of the quiver is reproduced by the intersection pairing of
the bound states of wrapped branes. This information along with the grading of
the boundary states by the characters lead to the identification of the
boundary states as bound states of wrapped branes on the resolution of the
orbifold.
\section{D-branes on the blow up}\label{gkzsec}
In the previous section we have obtained the boundary states in the CFT
describing fractional branes on the orbifold 
$\C^3/\Z_5$. 
In order to interpret these states geometrically, we have to describe
BPS branes wrapped on the exceptional divisors of the resolution
$\mathbf{X}$ in the large volume limit and show the emergence of the 
fractional branes as the cycles shrink at the orbifold. 
In other words, we need to find
the behaviour of the BPS-spectrum of the orbifold theory
under the marginal deformations
which parametrise the K\"ahler moduli space of $\mathbf{X}$. 
This, added to the fact that the boundary states are in one-to-one
correspondence with the irreducible representations of $G=\Z_5$, furnish a
physical realisation of the McKay correspondence \cite{reid,ito,craw1,craw2}. 
However, as mentioned in \S\ref{intro}, in order to avoid dealing with
world-sheet instanton corrections to the K\"ahler structure moduli space, we
shall go over to the complex structure moduli space of the mirror geometry.  
The latter follows from the periods of a meromorphic one-form on the 
mirror Riemann surface. The periods solve a set of differential equation 
known as the GKZ system of  equations
or the  generalised hypergeometric system. The solutions to the GKZ system
are also known as semi-periods. We shall use the expressions semi-period and
period interchangeably in this article.

In order to set up the notations let us recall
a few notions in toric geometry which are relevant for our purpose 
\cite{ful,katz1,katz2,ckyz,hosono1}. An $n$-dimensional toric
variety $\cal V$ is represented combinatorially as a fan $\Sigma$ in a real 
vector space $\bn_\R$ associated with a lattice $\bn$ which consists 
of a set of strongly convex rational polyhedral cones $\sigma$. 
A fan is described by a
set of $k$ points ($k>n$) $\{{\vec v}_i\}$ (which we shall refer to as the 
toric data) in $\bn$ and the cones are given upon its triangulation.
The convex hull of these points is called the Newton polytope $\triangle_N$.
The points satisfy a set of  $(k-n)$ linear relations $\{\ell_i^a\}$, \eg 
$\sum_i l^a_i {\vec v}_i = 0$.  Each of the points ${\vec v}_i$ corresponds 
to a homogeneous variable $x_i$ of $\C^k$, while a relation specifies a 
$\C^{\star}$ action on the homogeneous coordinates. This yields the
the $n$-dimensional toric variety as a quotient 
${\cal V} = (\C^k\setminus U)/(C^{\star})^{k-n}$.  $U$ is the set of 
singular points of the $\C^{\star}$ actions which is determined by 
the triangulation of the fan.

To every fan in the lattice $\bn_\R$  is  associated  a dual fan consisting of
cones in the dual space $\bm_\R$ where $\bm = \hom{\bn,\Z}$ is 
the dual lattice.  The cone $\dual{\sigma}$  dual to  
$\sigma$ is  given by $\dual{\sigma} = \{u\in\bm_{\R} |
\langle u,v \rangle \geq 0, \forall v\in\sigma\}$. 
The dual of the Newton polytope in $\bm_{\R}$ is given as
$\nabla_N = \{u\in \bm_\R|\;\langle u,v\rangle \geq -1,
\forall v\in\triangle_N\}$. The toric data 
associated with the dual fan are the points representing the vertices of this 
dual polytope. The dual polytope defines a  variety dual to $\cal V$ similarly
as above and for a large variety of cases these two varieties are related to
each other by mirror symmetry. 

In a fan associated to an orbifold, the singularity is signalled by
the presence of a cone which does not have unit volume in units of
lattice-spacing. A singularity can be resolved by refining the 
triangulation of the fan. 

In this section, we shall first briefly describe the toric 
resolution of the orbifold $\C^3/\Z_5$, followed by the triple-intersection
numbers of the exceptional divisors. 
We then discuss the local mirror geometry, which is a Riemann
surface of genus two, given as a polynomial equation of degree 5.
The GKZ equations for the mirror geometry are then derived from the toric
data. Finally, we obtain the semi-periods as solutions of the
GKZ system in the {\sc lcsl}.
\subsection{Resolution of $\C^3/\Z_5$ by D-branes}
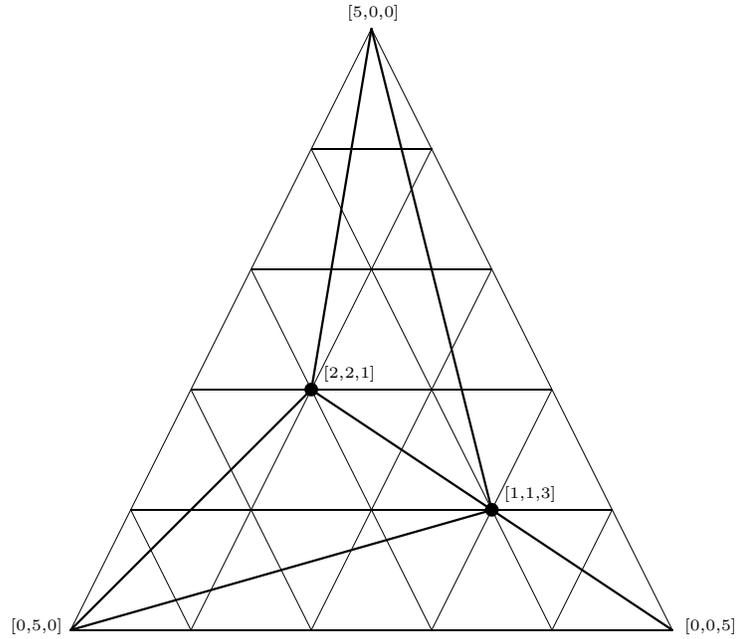
\begin{figure}[h]
\begin{center}
\setlength{\unitlength}{1.6mm}
\begin{picture}(50,50)(0,0)
\thinlines
\put(0,0){\line(1,2){25}}
\put(0,0){\line(1,0){50}}
\put(50,0){\line(-1,2){25}}
\put(5,10){\line(1,0){40}}
\put(10,20){\line(1,0){30}}
\put(15,30){\line(1,0){20}}
\put(20,40){\line(1,0){10}}
\put(10,0){\line(1,2){20}}
\put(20,0){\line(1,2){15}}
\put(30,0){\line(1,2){10}}
\put(40,0){\line(1,2){5}}
\put(10,0){\line(-1,2){5}}
\put(20,0){\line(-1,2){10}}
\put(30,0){\line(-1,2){15}}
\put(40,0){\line(-1,2){20}}
\thicklines
\put(0,0){\line(1,1){20}}
\put(0,0){\line(7,2){35}}
\put(50,0){\line(-3,2){30}}
\put(25,50){\line(-1,-6){5}}
\put(25,50){\line(1,-4){10}}
\put(20,20){\circle*{1}}
\put(35,10){\circle*{1}}
\put(-5,0){$\scriptscriptstyle [0,5,0]$}
\put(51,0){$\scriptscriptstyle [0,0,5]$}
\put(23,51){$\scriptscriptstyle [5,0,0]$}
\put(21,21){$\scriptscriptstyle [2,2,1]$}
\put(36,11){$\scriptscriptstyle [1,1,3]$}
\end{picture}
\end{center}
\caption{{\small\sl The junior simplex for $\frac{1}{5}[113]$}}
\label{jsmplxfig}
\end{figure}

A D0-brane on the blown up orbifold
$\C^3/\Z_5$ is obtained by starting with the supersymmetric linear sigma
model describing the world-volume gauge theory
of five coalesced D0-branes and then identifying these
under the $\Z_5$ symmetry. This is effected by an action of $\Z_5$ on the
Chan-Paton matrices of the open strings ending on the branes by the regular
representation of $\Z_5$. The vacuum moduli space of the world-volume gauge
theory, obtained by solving the $F$- and $D$-flatness conditions, provides the
resolution $\mathbf{X}$ of $\C^3/\Z_5$. The resolution $\mathbf{X}$ 
is described as a toric variety  by the charges of the chiral
multiplets in the linear sigma model \cite{dgm}, and is given as the following
toric data 
\begin{eqnarray}
\label{tordat}
{\cal A} =
\bordermatrix{
&\D_0&\D_3&\D_4&\D_1&\D_2&\nonumber \\
&0&1&0&0&-1\nonumber \\
&0&0&1&-1&-3\nonumber \\
&1&1&1&1&1\nonumber \\
}
\end{eqnarray}
The five columns of the matrix $\cal A$  define a set of five integral 
points on the height-one plane of the third coordinate in
the lattice $\mathbf N\cong \Z^3\subset\R^3$. 
\begin{figure}[h]
\begin{center}
\setlength{\unitlength}{1.3cm}
\thicklines
\begin{picture}(5,5)(-3,-3)
\put(0,0){\circle*{.1}}
\put(1,0){\circle*{.1}}
\put(0,1){\circle*{.1}}
\put(0,-1){\circle*{.1}}
\put(-1,-3){\circle*{.1}}
\put(-1,-3){\line(2,3){2}}
\put(-1,-3){\line(1,4){1}}
\put(1,0){\line(-1,1){1}}
\put(.1,0){$\scriptscriptstyle (0,0)$}
\put(1.2,0){$\scriptscriptstyle (1,0)$}
\put(-.2,1.2){$\scriptscriptstyle (0,1)$}
\put(-1.9,-3){$\scriptscriptstyle (-1,-3)$}
\put(-.2,-.8){$\scriptscriptstyle (0,-1)$}
\end{picture}
\end{center}
\caption{{\small\sl Plot of the columns of the toric data ${\cal A}$
on the height-one plane --- the Newton polygon $\Delta$}}
\label{figz5}
\end{figure}
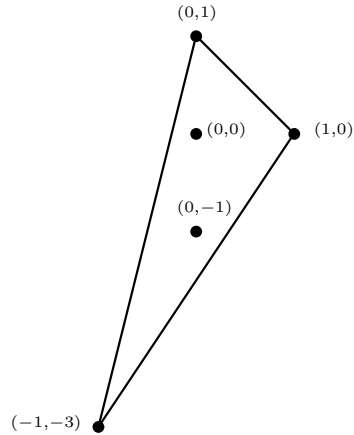
The convex hull
of these points in $\R^3$ is the Newton polytope. Its trace on 
the height-one plane is the Newton polygon $\Delta$,  
as shown in Figure~\ref{figz5}. 
We have denoted the toric divisors corresponding to the points in the Newton
polygon by $\goth{D}$. From the toric data $\cal A$
we see that the toric divisors are linearly related to each other, 
\begin{equation}
\label{reln}
\begin{split}
{\D}_2 &={\D}_3 \stackrel{\mathrm{def}}{=}\goth{d}_1\\
{\D}_4&\stackrel{\mathrm{def}}{=}\goth{d}_2\\ 
{\D}_0&=\goth{d}_1-2\goth{d}_2\\
{\D}_1&=\goth{d}_2-3\goth{d}_1,
\end{split}
\end{equation} 
where we introduced the two linearly independent divisors $\goth{d}_1$ and
$\goth{d}_2$, which can be taken to be the generators of the Picard group,
$\pic{\mathbf{X}}=\Z\oplus\Z$.
\begin{figure}[h]
\begin{center}
\setlength{\unitlength}{1.3cm}
\begin{picture}(5,5)(-3,-3)
\put(0,0){\circle*{.1}}
\put(1,0){\circle*{.1}}
\put(0,1){\circle*{.1}}
\put(0,-1){\circle*{.1}}
\put(-1,-3){\circle*{.1}}
\thicklines
\put(-1,-3){\line(2,3){2}}
\put(-1,-3){\line(1,4){1}}
\put(1,0){\line(-1,1){1}}
\thinlines
\put(-1,-3){\line(1,2){1}}
\put(-1,-3){\line(1,3){1}}
\put(1,0){\line(-1,0){1}}
\put(1,0){\line(-1,-1){1}}
\put(0,1){\line(0,-1){2}}
\put(.1,.1){$\scriptscriptstyle {\D}_0$}
\put(.05,-.7){$\scriptscriptstyle {\D}_1$}
\put(-1.4,-3){$\scriptscriptstyle {\D}_2$}
\put(1.2,0){$\scriptscriptstyle {\D}_3$}
\put(-.1,1.1){$\scriptscriptstyle {\D}_4$}
\end{picture}
\end{center}
\caption{{\small\sl Triangulation of the Newton polygon}}
\label{trgz5}
\end{figure}
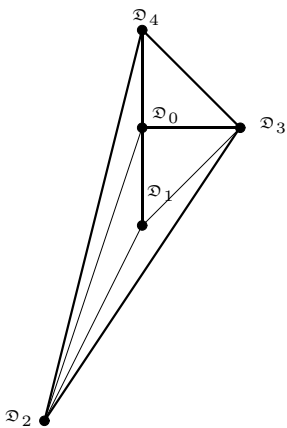
The resolution $\mathbf{X}$ of the orbifold corresponds to 
triangulating on all the points of the Newton polygon, as shown in
Figure~\ref{trgz5}. 
Let us point out that
the same triangulation of the Newton polygon can also be achieved, playfully,
by observing the resemblance between the triangulation in 
Figure~\ref{trgz5} and the junior simplex \cite{ito,reid} corresponding to 
the orbifold, by the survivours of the \emph{champions' meet}
\cite{craw1,craw2}, as in Figure~\ref{jsmplxfig}.
The Newton polygon is not reflexive, as it has two
points in its interior implying $h^{11}=2$ for the resolved orbifold. 
The divisors $\goth{D}_0$ and $\goth{D}_1$, corresponding
to the internal points are the exceptional divisors. As shown in 
Figure~\ref{trgz5} and Figure~\ref{jsmplxfig}, the two
interior points corresponding to $\goth{D}_0$ and $\goth{D}_1$
have line-valency 4 and 3, corresponding, thereby, to a
rational scroll $\mathbb{F}_2$, which is a $\cp{1}$-bundle over $\cp{1}$ 
and the projective plane $\cp{2}$, respectively \cite{reid,craw1,craw2}. 

The Gale diagram of the $3\times 4$ matrix $\cal A$, assuming  
value in the kernel of $\cal A$, is 
\begin{equation}
{\cal G}=\begin{pmatrix}
1&-2\\1&0\\0&1\\-3&1\\1&0
\end{pmatrix}\in\ker{\cal A}.
\end{equation}
To  the five three-vectors of the Newton polyhedron 
is associated a lattice of linear
relations, generated by the columns of the Gale diagram.
Let us  denote it by ${\cal L}_{\cal A}$; it is a
sublattice of $\Z^5$. The canonical bundle of
$\mathbf{X}$ is trivial, as may be seen from the fact that
the sum of the components of each element
$\ell$ of ${\cal L}_{\cal A}$ vanishes, 
\begin{equation} 
\sum_{i=0}^4\ell_i=0, \quad \ell\in {\cal L}_{\cal A},
\end{equation} 
signifying that the resolution is crepant.
From the relations  
\begin{equation}
\label{kerlcsl}
\begin{matrix}
\ell^{(1)} &=&( 1&1&0&-3&1)\\
\ell^{(2)} &=&( -2&0&1& 1&0)\\
\end{matrix}
\end{equation}
we can write down the toric ideal of $\mathbf{X}$, namely,
$\langle y_0y_1y_4 - y_3^3,\, y_2y_3-y_0^2\rangle$, where we have written the
generators of the ideal in Gr\"obner basis. Consequently, 
the relations $\ell^I$ generate the Mori cone of the variety associated to
$\Delta$.
\subsection{Yukawa couplings}
\label{yukaba}
In order to study the B-type branes wrapped on the cycles corresponding to
the meromorphic
form on $\dual{\mathbf{X}}$, we need to derive the periods as solutions to the
GKZ system. To write the solutions in the {\sc lcsl} in the canonical basis 
we need to know the triple-intersection numbers of the divisors, alias 
Yukawa couplings.
Let us consider the triangulation of the Newton polygon 
shown in Figure~\ref{trgz5}.
Denoting the intersection number 
$\goth{D}_i\cap\goth{D}_j\cap\goth{D}_k$
by $C_{ijk}$, 
we have the following values for Yukawa couplings \cite{hkty1,hkty2}, 
corresponding to the two divisors $\mathfrak{D}_0$ and $\mathfrak{D}_1$,
\begin{equation}
\begin{split}
{C}_{000} = 8 &\qquad C_{001}=1\\
C_{011}=-3 &\qquad C_{111}=9,
\end{split}
\end{equation}
the other couplings vanish.
Using the relations \eq{reln} we can derive the triple intersection numbers 
corresponding to the divisors $\goth{d}$, 
\begin{equation}
\begin{split}
c_{111}= -\frac{2}{5} &\qquad c_{112}= -\frac{1}{5}\\
c_{122}= -\frac{3}{5} &\qquad c_{222}= -\frac{9}{5},
\end{split}
\end{equation}
where now $c_{ijk}$ denotes the intersection number of 
$\goth{d}_i\cap\goth{d}_j\cap\goth{d}_k$.
In the next section we shall use the couplings $c_{ijk}$ to obtain the
solutions to the GKZ system \eq{gkzop} in the {\sc lcsl}. 
\subsection{The mirror}
As mentioned earlier, in order to obtain the exact K\"ahler moduli space of the 
orbifold we need to consider the moduli space of complex structure of 
the mirror 
geometry. The construction of local mirror
$\dual{\mathbf{X}}$ of $\mathbf{X}$ follows by making use of the fact that a 
pair of dual Newton polygons in dual lattice spaces $N$ and $M$ defines a pair 
of mirror varieties. 
We shall write the mirror of the non-compact variety as a
curve \cite{katz1,katz2,ckyz,hosono1,hiv} defined by the polynomial equation
\begin{equation}
\sum_{i=0}^4 b_iy_i = 0,
\end{equation}
where the $y_i$ appeared in the binomials of the generators of the toric
ideal and the $b_i$ are five complex numbers. 
The coefficients of the monomials represent the deformation of complex structure 
or equivalently the blow up of the K\"ahler structure of the mirror geometry. 
However, all of them are not independent and the appropriate choice is 
associated with the choice of the kernel. In order to make use of the mirror map 
one has to choose the 
Mori basis of kernel so that the independent coordinates of complex structure 
moduli space  of mirror geometry correspond to the generators of the 
K\"ahler cone 
in the moduli space of K\"ahler structure in the blown up
geometry. The moduli space is singular in general. The singular set is given by 
the principal component  of the discriminant. 

We can rewrite the equation in terms of homogeneous variables $\tilde y_i$
as 
\begin{equation}
P(\tilde y_1,\tilde y_2,\tilde y_3)\thinspace{\equiv}\thinspace
b_1 \tilde y_1^5 + b_4\tilde y_2^5 + b_2 \tilde y_3^5 + b_0 \tilde y_1
\tilde y_2\tilde y_3^3 
+ b_3 \tilde y_1^2 \tilde y_2^2 \tilde y_3=0. 
\label{poly1}
\end{equation}
Let us note at this point the
usefulness of the junior simplex in obtaining the homogeneous variables. The
indices of the homogeneous variables in the 
monomials in the above equation are given by  the points marked in
Figure~\ref{jsmplxfig} that participate in the triangulation.  We then 
rescale the coordinates $\tilde y_i$, using the  $(\cstar)^3$ symmetry 
\begin{equation}
\begin{split}
\tilde y_1 &= x_1 b_1^{-1/5}, \\
\tilde y_2 &= x_2 b_4^{-1/5}, \\
\tilde y_3 &= x_3 b_2^{-1/5}. \label{scale1}
\end{split}
\end{equation}
With this substitution the equation \eq{poly1} yields a curve of degree five,
with two complex deformation parameters, 
\begin{equation}
\label{mirroreq}
P(y_1,y_2,y_3)
 = x_1^5 + x_2^5 + x_3^5 +\zeta_1^{1/5} x_1x_2x_3^3 + 
\zeta_2^{1/5} x_1^2x_2^2x_3,
\end{equation}
where 
\begin{equation}
\zeta_1 = \frac{b_0^5}{b_1b_4b_2^3} 
\qquad\qquad 
\zeta_2 = \frac{b_3^5}{b_1^2b_4^2b_2}.
\label{para1}\end{equation}
We also define $z_1=1/\zeta_1$ and $z_2=1/\zeta_2$.
Let us note that $(\zeta_1,\zeta_2)\rt (0,0)$ yield  the orbifold limit
in the mirror,
while the {\sc lcsl} is at the small values of $z_1$, $z_2$.

Since the large K\"ahler structure limit of $\mathbf{X}$ corresponds to the
{\sc lcsl} of the mirror $\dual{\mathbf{X}}$, 
the A-type branes in the large volume limit correspond to B-type branes in the 
{\sc lcsl} in the mirror. The coordinate chart suitable to the {\sc lcsl} in
$\dual{\mathbf{X}}$ is given by combinations of the parameters $b_i$, with
their indices in the combinations determined by the relations which correspond
to the Gr\"obner basis of the toric ideal, as mentioned above. 
Thus, the good coordinates in the {\sc lcsl} are
\begin{equation}
u=-\frac{b_0b_1b_4}{b_3^3} \qquad v=\frac{b_2b_3}{b_0^2},
\end{equation}
corresponding to the relations $\ell^{(I)}$ in \eq{kerlcsl}.
These are related to the $z_1,z_2$ defined above by 
\begin{equation}
z_1=uv^3 \qquad z_2=u^2v.
\end{equation}
The variables $z_1$ and $z_2$, as seen from \eq{para1}, 
correspond to the relations 
\begin{equation}
\label{kerorb}
\begin{matrix}
L^{(1)}=& (-5 & 1 & 3 & 0&1)\\
L^{(2)}=& (0 & 2& 1&-5&2)
\end{matrix}
\end{equation} 
in the same way $u$,$v$ correspond to \eq{kerlcsl}. The corresponding toric
ideal is not Gr\"obner.

The discriminant locus of the curve \eq{mirroreq} has three distinct
components. The {\sc lcsl}, given by  $(u,v)=(0,0)$,  the orbifold point, given by
$(\zeta_1,\zeta_2)=(0,0)$ and finally, a curve, namely, 
the principal component, given by
\begin{equation}
\label{PD}
(1-4v)^2+3125u^2v^3-u(27-225v+500v^2)=0 .
\end{equation}
\subsection{The GKZ system}
According to the local mirror conjecture the volume of the holomorphic cycles 
are related to the periods of a meromorphic one-form on the mirror geometry, the 
Riemann surface. In general it is not possible to integrate explicitly the 
one-form on the cycles of the Riemann surface. For a toric description there is 
a technical simplification due to the fact that all the periods satisfy a set of 
differential equations which one can directly obtain from the toric data. They 
are in general called $\cal A$-hypergeometric system and we are going construct 
the system for the present case. 

Considering the affine space $\C^5$ with
coordinates $b=\{b_0,{\ldots}, b_4\}$, for each $\ell\in{\cal L}_{\cal A}$, 
we define the homogeneous differential operator \cite{gkz,ode,hly}
\begin{equation}
{\cal D}_{\ell}=\prod_{\ell_i>0}\left[\frac{\pa}{\pa b_i}\right]^{\ell_i}
-\prod_{\ell_i<0}\left[\frac{\pa}{\pa b_i}\right]^{-\ell_i}.
\end{equation}
The set of solutions $\Phi(b)$ to the system of equations 
\begin{equation}\label{gkzsys} 
{\cal D}_{\ell^{(I)}}\Phi(b)=0,
\end{equation}
corresponding to a basis of ${\cal L}_{\cal A}$ furnish the periods in some
region of the moduli space of complex structure deformations.

For the toric data $\cal A $,
the periods in {\sc lcsl} are given by the kernels of the GKZ system of partial
differential operators, written in terms of the variables $u$,$v$.
Since $\mathsf{vol}({\cal A})=5$, as can be verified from Figure~\ref{trgz5}, 
there are five linearly independent
solutions to the GKZ system.
These operators corresponds to the basis of the lattice of relations are
\begin{equation}
\begin{split}\label{gkzop}
{\cal D}_1 & = (\t{u}-2\t{v})\t{u}^2- u(\t{v}-3\t{u}) (\t{v}-3\t{u}-1)
(\t{v}-3\t{u}-2),\\
{\cal D}_2 & = (\t{v}-3\t{u})\t{v}- v(\t{u}-2\t{v}) (\t{u}-2\t{v}-1)
\end{split}
\end{equation}
where $\t{x}=x\frac{d}{dx}$ denotes the logarithmic derivative 
with respect to $x$.

Let us also record the GKZ operators in terms of the variable $z_1$,$z_2$.
\begin{equation}
\begin{split}
\dual{\cal D}_1 =(2\t2+\t1)^2 (3\t1+\t2) (2\t2&+\t1-1)^2 
\\&+ z_2 \t2(\t2+\frac{1}{5})(\t2+\frac{2}{5})(\t2+\frac{3}{5})
(\t2+\frac{4}{5}),  \\
\dual{\cal D}_2=(2\t2+\t1)^2 (3\t1+\t2) (3\t1&+\t2 -1) (3\t1 + \t2-2)\\ 
&+  z_1 \t1(\t1+\frac{1}{5})(\t1+\frac{2}{5})
(\t1+\frac{3}{5})(\t1+\frac{4}{5}). 
\end{split}
\end{equation}
where $\t{i}=z_i\frac{d}{dz_i}$ for $i=1,2$.
Both the equations being of fifth order, this system also has five independent
solutions.
\section{In the large complex structure limit}
\label{lcslsec}
In this section we shall obtain all the periods near the {\sc lcsl}. 
The periods on the cycles in $\mathbf{X}$  for the case at hand 
are in one-to-one correspondence with the solutions to the GKZ-system.
The $n$-cycle periods can be identified, up to subleading terms, with the 
solutions to the GKZ-system whose leading terms are $n^{\mathrm{th}}$ 
powers (or their combinations) of the logarithm of the moduli space variables.
A complete identification, however, is not straightforward. One way to fix 
that is through the monodromies around all the singularities 
of the moduli space. In the present example, however, the  
principal component of discriminant locus is a complicated curve. 
Hence we shall not calculate the monodromy around that. We shall make 
a few simplifying assumptions to obtain all the solutions. 

We start by finding out the solutions to the GKZ system using  Frobenius'
method around the point $(u,v)=(0,0)$. These correspond to the periods of
cycles of the exceptional divisor of the blow up in 
the {\sc lcsl}. In terms of the differential equations, the {\sc lcsl}, $(u,v)=(0,0)$,  
is characterised by the maximal degeneration of the solutions to the indicial
equations associated to ${\cal D}_1$ and ${\cal D}_2$, which correspond to
the Mori basis of the relations, $\ell^{I}$ \cite{maxim}.
More specifically, all the indices are zero.
Therefore, in this limit, the equations have 
a single power series solution,
two solutions containing logarithms of $u$ and $v$, as
well as two solutions containing squares of the logarithms. 

The power series solution, also called the fundamental period, 
can be obtained  from the relation $\ell^{(I)}$. 
Let us write it as 
\begin{equation} 
\varpi_{0} = \sum_{m,n=0}^{\infty} a(m,n) u^mv^n.
\end{equation}
The coefficients $a(m,n)$  of the series satisfy the recursion relations
\begin{equation}
\begin{split}
\frac{a(m+1,n)}{a(m,n)}&=\frac{
(n-3m)(n-3m-1)(n-3m-2)}{(m-2n+1)(m+1)},\\
\frac{a(m,n+1)}{a(m,n)}&=\frac{(m-2n)(m-2n-1)}{(n-3m+1)(n+1)},
\end{split}
\end{equation}
derived from \eq{gkzop}.
From these relations we can find out the associated radii of convergence of
the series as the upper bounds on $u$, $v$. The domain of convergence is 
bounded by $|u|<u_0$ and $|v|<v_0$.
The associated radii of convergence $u_0$ and $v_0$ trace out a curve in the
$u$--$v$ plane. The curve is parametrised by $x$ and $y$, 
defined through
\begin{equation} 
u_0 = \frac{x^2 (x-2y)}{(y-3x)^3},\qquad v_0 = \frac{y (y-3x)}{(x-2y)^2}
\end{equation}
and is given by the following equation in terms of $u_0$ and $v_0$
\begin{equation}
(1-4v_0)^2+3125u_0^2v_0^3-u_0(27-225v_0+500v_0^2)=0.
\end{equation}
Let us note that this is the same as the principal component of the
discriminant locus \eq{PD} with $u$ and $v$ replaced by $u_0$ and $v_0$,
respectively. From this equation, we also find that $\mathsf{max}[u_0] =
\frac{1}{27}$ and $\mathsf{max}[v_0]=\frac{1}{4}$. The coefficient $a(m,n)$ of
$\varpi_0$ in this domain, which solves the recursion relations and
conforms to the toric data, is given by   
\begin{equation}
a(m,n) = \frac{1}{\Gm{1+m}^2\Gm{1+n}\Gm{1+m-2n}\Gm{1-3m+n}}.
\end{equation}
Let us note that with this coefficient, the only contribution to
$\varpi_0(u,v)$ is from the term $m=n=0$, since all the other terms have
poles of the Gamma functions in the denominator of the series. Hence 
$\varpi_0=1$.

The other solutions to the GKZ system can be obtained from $\varpi_0$ by
taking indicial derivatives. The two solutions linear in the logarithms of $u$
and $v$ are given by 
\begin{equation}\label{log1}
\begin{split}
\varpi_1 &=\lim_{\rho\rt 0}\frac{1}{2\pi i}
\pa_{\rho}\left[\sum_{m,n=0}^{\infty}
a(m+\rho ,n)u^{m+\rho}v^n\right], \\
\varpi_2 &=\lim_{\rho\rt 0}\frac{1}{2\pi i}
\pa_{\rho}\left[\sum_{m,n=0}^{\infty}
a(m,n+\rho)u^mv^{n+\rho} \right]. 
\end{split}
\end{equation}
The two solutions quadratic in the logarithms, on the other hand, are obtained
by taking linear combinations of the second derivatives with respect to the 
indices, with coefficients determined by the  Yukawa couplings as
\cite{hkty1,hkty2}
\begin{equation}\label{log2}
\begin{split}
\varpi_3 &= \lim_{\rho_i\rt 0} 
\frac{1}{2!}\frac{1}{(2i\pi)^2}
\sum_{j,k=1}^2 c_{2jk}\pa_{\rho_j}\pa_{\rho_k}
\left[\sum_{m,n=0}^{\infty}
a(m+\rho_1 ,n+\rho_2)u^{m+\rho_j}v^{n+\rho_k}\right],\\ 
\varpi_4 &=\lim_{\rho_i\rt 0} 
\frac{1}{2!}\frac{1}{(2i\pi)^2}
\sum_{j,k=1}^2 c_{1jk}\pa_{\rho_j}\pa_{\rho_k}
\left[\sum_{m,n=0}^{\infty}
a(m+\rho_1 ,n+\rho_2)u^{m+\rho_j}v^{n+\rho_k}\right]. 
\end{split}
\end{equation}

Using the definitions \eq{log1} and \eq{log2}, let us now write down the
explicit expressions of the solutions.
Unlike $\varpi_0$, in all these four solutions the series receive
contributions from non-vanishing $m$ and $n$. 
We have 
\begin{equation}
\begin{split} \label{varpi1}
\varpi_1(u,v) = \frac{\varpi_0}{2\pi i}\log u  
+&\frac{3}{2\pi i}\sum_{(n,r)'}
\frac{\Gm{5n+3r}}{\Gm{1+r}\Gm{1+n}\Gm{1+2n+r}^2}(-u^2v)^n (-u)^r \\
-&\frac{1}{2\pi i}\sum_{(m,r)'}
\frac{\Gm{5m+2r}}{\Gm{1+r}\Gm{1+3m+r}\Gm{1+m}^2}(-uv^3)^mv^r ,
\end{split}
\end{equation}
where we used an abbreviation: 
$(m,n)'\equiv (m,n)\in\{\Z_{+}^2\setminus {(0,0)}\}$. The other period linear
in the logarithm is  
\begin{equation}
\begin{split} \label{varpi2}
\varpi_2(u,v) 
= \frac{\varpi_{0}}{2\pi i}\log v 
&-\frac{1}{2\pi i}\sum_{(n,r)'}
\frac{\Gm{5n+3r}}{\Gm{1+r}\Gm{1+n}\Gm{1+2n+r}^2}(-u^2v)^n (-u)^r \\
&+\frac{2}{2\pi i}\sum_{(m,r)'}
\frac{\Gm{5m+2r}}{\Gm{1+r}\Gm{1+3m+r}\Gm{1+m}^2}(-uv^3)^mv^r .
\end{split}
\end{equation}
While the solutions are written in the chart $(u,v)$, in the neighbourhood of 
$(0,0)$,
each solution splits into terms whose ``natural" variables are $(z_1,v)$ and
$(u,z_2)$. We may, thus, take linear combinations of $\varpi_{1}$ and
$\varpi_{2}$, which are expressed in these mixed set of variables. 
The combination
\begin{equation}\label{Pi1}
\begin{split}
\Pi_1 &= \varpi_1 + 3\varpi_2 \\
&= 
\frac{1}{2\pi i}\log z_1 + \frac{5}{2\pi i}
\sum_{(m,r)'}
\frac{\Gm{5m+2r}}{\Gm{1+r}\Gm{1+3m+r}\Gm{1+m}^2}(-z_1)^mv^r, 
\end{split}
\end{equation}
is expressed in terms of $z_1$ and $v$, while the combination 
\begin{equation}\label{Pi2}
\begin{split}
\Pi_2 &= \varpi_2 + 2\varpi_1 \\
&= \frac{1}{2\pi i}\log z_2 + \frac{5}{2\pi i}
\sum_{(n,r)'}
\frac{\Gm{5n+3r}}{\Gm{1+r}\Gm{1+2n+r}^2\Gm{1+n}}(-z_2)^n (-u)^r 
\end{split}
\end{equation}
is expressed in terms of $u$ and $z_2$.
The solutions quadratic in the logarithms can also be found similarly. One of
them, $\varpi_3$ takes the following form,
\begin{equation}
\begin{split}
\varpi_3 = &-\frac{1}{10}\frac{1}{(2\pi i)^2} \left[ 2 (\varpi_1+3\varpi_2)
\log z_1 - \varpi_0(\log z_1 )^2 - 36 \Psi'(1)\right]\\
&-\frac{1}{(2\pi i)^2}\sum_{(m,r)'}\frac{5\Psi(5m+2r)-2\Psi(1+m)-3\Psi(1+3m+r)}
{\Gm{1+r}\Gm{1+m}^2\Gm{1+3m+r}} \Gm{5m+2r} (-z_1)^m v^r.
\end{split} 
\end{equation} 

The two solutions linear in logarithms represent the two 
semi-periods which are identified with the K\"ahler moduli
using local mirror symmetry 
\cite{hkty1,hkty2,ckyz}. 
According to the principle of local mirror symmetry, 
the flat coordinates $t_1$ and $t_2$ in the {\sc lcsl} have the 
asymptotic behaviours \cite{aspin}
\begin{equation}
t_1 = \frac{1}{2\pi i}\log(- u) +  {\cal O}(u,v),
\qquad
t_2 = \frac{1}{2\pi i}\log(v)  + {\cal O}(u,v),
\end{equation}
which imply the following identifications
\begin{equation}
t_1 = \varpi_1 + \frac{1}{2},
\qquad
t_2 = \varpi_2 
\end{equation}
between the flat coordinates and the periods.

The other two semi-periods dual to these are not exactly equal to the 
$\varpi_3$, $\varpi_4$. There is  an ambiguity of adding linear 
combinations of $\varpi_1$ and $\varpi_2$ to these.
In order to fix the periods dual to $t_1$ and $t_2$ 
we demand that in the {\sc lcsl} they asymptote to
\begin{equation}
\label{tds}
\begin{split}
t_{d1}&=\frac{1}{2}\sum_{i,j} c_{1ij}{t_i}{t_j} + m_1 +{\cal O}(u,v),
\\
t_{d2}&=\frac{1}{2}\sum_{i,j} c_{2ij}{t_i}{t_j} + m_2 + {\cal O}(u,v),
\end{split}
\end{equation}
where $m_1$ and $m_2$ are two constants, interpreted as the mass of a 
D4-brane wrapped on the cycle.
We shall denote the solutions that asymptote to $t_{d1}$ and
$t_{d2}$ as $\Pi_3$ and $\Pi_4$, respectively. This fixes the integral 
symplectic transformation that shifts the $t_{di}$ by 
the $t_i$. The constants may be fixed by 
demanding that they should vanish on the principal component of the 
discriminant locus. We shall fix these later from the equality of the masses
of the fractional branes.
While discussing the identification of the brane configurations 
we shall find this to be the natural choice.

With the above choice of asymptotics we can now write down the solutions  
quadratic in logarithms as follows. One of them,
$\Pi_3$, is naturally expressed in terms of the mixed variables as
\begin{equation}
\begin{split}
\Pi_3(z_1,z_2) 
= &-\frac{1}{10} \frac{1}{2\pi i}
\left[\log^2 (-z_1) 
+ \frac{10}{2\pi i}\log (-z_1)
\sum_{(m,r)'}
\frac{\Gm{5m+2r}}{\Gm{1+r}\Gm{1+3m+r}\Gm{1+m}^2}(-z_1)^mv^r\right] \\
&-\frac{1}{(2\pi i)^2}\sum_{(m,r)'}\frac{5\Psi(5m+2r)-2\Psi(1+m)-3\Psi(1+3m+r)}
{\Gm{1+r}\Gm{1+m}^2\Gm{1+3m+r}} \Gm{5m+2r} (-z_1)^m v^r\\ 
  &+\frac{1}{10}\frac{18\Psi'(1)}{(2\pi i)^2}.
\end{split}
\end{equation}
The other solution quadratic in the logarithms, $\Pi_4$,
can be succinctly expressed through the combination $\Pi_3-3\Pi_4 $ as
\begin{equation}
\Pi_3 -3\Pi_4 = \frac{1}{2}\frac{1}{(2\pi i)^2} \log^2 (-u) + \frac{1}{2\pi 
i}\log (-u) \varpi_1 - \frac{12\psi'(1)}{2(2\pi i)^2}.
\end{equation}
It may be noted that $\Pi_3$ and $\Pi_4$ are linear combinations of $\varpi_4$ 
and $\varpi_3$, respectively, with the other two semi-periods.

The monodromy around the point $(u,v)=(0,0)$  
can be obtained from the above formulas by expressing the changes of the
solutions while going around the loops $(u,v)\rt (e^{2i\pi}u, v)$ 
and $(u,v)\rt (u, e^{2i\pi}v)$ in terms of the original solutions. 

We shall find shortly that expressions of the solutions in terms of the mixed
variables are easier for analytic continuation to the orbifold region. The
second solution quadratic in logarithms, namely $\Pi_4$, does not admit
a similar form, thus rendering its continuation to the orbifold region
difficult.
\section{At the orbifold point}
\label{orbsec}
The semi-periods obtained in the last section correspond to branes wrapped on
the cycles at large volume. The fractional branes pertain to the region of the
moduli space where the exceptional 
cycles shrink, giving way to the orbifold singularity.
Thus, in order to compare the wrapped branes and the fractional branes we have
to relate the periods in these two regions. One way of doing this is by 
analytic continuation of the large volume periods to the orbifold region. 
In the local mirror description \eq{mirroreq}, the orbifold point 
is located at $(\zeta_1, \zeta_2)\rt (0,0)$, which, in turn, 
corresponds to the point $(u, v)\rt(\infty,\infty)$.
In this section we shall first perform the analytic continuation of the
periods from the {\sc lcsl} to the orbifold point. We shall then obtain the
monodromy transformation of the periods and thence obtain the bound states of
D4-, D2- and D0-branes which are identified with the fractional branes as the
exceptional divisor shrinks.

\subsection{Analytic continuation of periods}
Analytic continuation of the periods is performed with the help of
Mellin-Barnes-type integral representations
of the periods obtained in the last section 
\cite{candel0,candel1,candel2,mayr}. 
Let us begin with
$\Pi_1$. We write it as a contour integral as
\begin{equation}
\Pi_1 = \frac{1}{(2\pi i)^2}\sum_{r=0}^{\infty}\frac{v^r}{\Gm{1+r}}
\oint\frac{\Gm{1+s}^2\Gm{-s}^2}
{\Gm{1+s}^2\Gm{1+3s+r}\Gm{1-5s-2r}}z_1^s,
\end{equation}
where the contour of integration encloses the poles of $\Gm{-s}$ at the
positive integral values of $s$.
For analytic continuation, we first rewrite this as 
\begin{equation}
\label{rerite}
\Pi_1= \sum_{r=0}^{\infty}\frac{v^r}{\Gm{1+r}}
\oint\frac{\Gm{5s+2r}\Gm{-s}}{\Gm{1+s}\Gm{1+3s+r}}\phi(s) z_1^s,
\end{equation}
where we defined 
\begin{equation}
\begin{split}
\label{phase}
\phi(x) &= \frac{\sin 5\pi x}{\sin\pi x} \\
&= 5- 20 \sin^2 x + 16 \sin^4 x.
\end{split}
\end{equation}
It can be checked that the last two terms in $\phi$ do not contribute to the
integral in \eq{rerite}. Thus, we can set $\phi$ to 5 
in \eq{rerite} and obtain,
\begin{equation}
\Pi_1= 5\sum_{r=0}^{\infty}\frac{v^r}{\Gm{1+r}}
\oint\frac{\Gm{5s+2r}\Gm{-s}}{\Gm{1+s}\Gm{1+3s+r}} z_1^s.
\end{equation}
The continued expressions are obtained on evaluating the integral after
deforming the contour of integration to enclose poles on the negative
real-axis in the $s$-plane. This yields
\begin{equation}
\Pi_1(\zeta_1,\zeta_2)=- \frac{5}{2\pi i} 
\sum_{(m,r)'}
\frac{\Gm{\frac{2r+m}{5}}{\zeta_1}^{\frac{m}{5}}{\zeta_2}^{\frac{r}{5}}}
{\Gm{1+r}\Gm{1+m}\Gm{1-\frac{2r+m}{5}}\Gm{1-\frac{3m+r}{5}}}
(-1)^m.
\end{equation}

The solution $\Pi_3$ quadratic in logarithms can be continued in a 
similar fashion by first writing it as 
\begin{equation}
\Pi_3 = -\frac{1}{5}\frac{1}{(2\pi i)^3} 
\sum_{r=0}^{\infty}\frac{v^r}{\Gm{1+r}}
\oint\frac{\Gm{1+s}^3\Gm{-s}^3}{\Gm{1+s}^2\Gm{1+3s+r}\Gm{1-5s-2r}} (-z_1)^s + 
m_2,
\end{equation}
where again the contour encloses poles on the positive real $s$-axis. 
In continuing the solutions quadratic in the logarithms, the $\sin^2 x$ term
in $\phi$ contributes to the integral by a constant. The contribution to
$\Pi_3$ is absorbed in the constant $m_2$, which will be left  arbitrary 
for the moment. We shall fix it later in this section.
There is again a similar expansion as in (\ref{phase}), and the constant coming 
from the second term while last term will not contribute. Then
using identities of Gamma function as before and deforming the contour 
we continue it to
\begin{equation}
\Pi_3(\zeta_1,\zeta_2)=\frac{1}{(2\pi i)^2} 
\sum_{(m,r)'}
\frac{\Gm{\frac{2r+m}{5}}^2
{\zeta_1}^{m/5}{\zeta_2}^{r/5}}
{\Gm{1+r}\Gm{1+m}\Gm{1-\frac{3m+r}{5}}}
(-1)^{\frac{4m-2r}{5}} + m_2.
\end{equation}

A similar consideration applies for $\Pi_2$.
Its continuation leads to the following expression.
\begin{equation}
\Pi_2(\zeta_1,\zeta_2)= -\frac{5}{2\pi i}\sum_{(m,r)'}
\frac{\Gm{\frac{3r+n}{5}}{\zeta_1}^{r/5}{\zeta_2}^{n/5}}
{\Gm{1+r}\Gm{1+n}\Gm{1-\frac{2n+r}{5}}^2}
(-1)^r(-1)^n
\end{equation}

From the expressions of $\Pi_1$ and $\Pi_2$ we can obtain the continuations of
the semi-periods $\varpi_1$ and $\varpi_2$ by inverting the relations \eq{Pi1}
and \eq{Pi2}.

The other semi-period $\Pi_4$ does not yield to such simple manoevres. 
A linear combination of the two semi-periods can also be written as a simple 
contour integral as
\begin{equation}
\Pi_3 - 3\Pi_4  =  - \frac{1}{(2\pi i)^4} \oint {ds}\oint{dt}
\frac{\Gm{1+s}{\Gm{-s}}^3\Gm{-t}}{\Gm{1+s-2t}\Gm{1-3s+t}} (-u)^s v^t 
+m_2-3m_1,
\end{equation}
but continuation to the orbifold is rather cumbersome. The
form of $\Pi_4$, however, can be obtained by demanding special properties 
of the monodromy transformation, as we shall discuss shortly. Here we 
quote the result thus obtained for completeness:
\begin{equation}
\Pi_4(\zeta_1,\zeta_2)=\frac{1}{(2\pi i)^2} 
\sum_{(m,r)'}
\frac{\Gm{\frac{2r+m}{5}}^2 \Gm{1-\frac{3m+r}{5}}
{\zeta_1}^{m/5}{\zeta_2}^{r/5}}
{\Gm{1+r}\Gm{1+m}}
(-1)^{\frac{2r+m}{5}} + m_1,
\end{equation}
where once again, $m_1$ is an arbitrary constant to be fixed.
\subsection{The monodromy} 
In order to identify the periods obtained above with the fractional brane
states of the CFT, we now have to study the behavior of the periods under
monodromy transformations around the orbifold point. The  basis of 
periods that correspond to the boundary states, by virtue of forming an
orbit of the monodromy matrix, can be obtained by taking the solutions around
the orbifold point in a loop. Let us define 
${\zeta'}_1$ and ${\zeta'}_2$, such that
$\zeta_1^{\prime 5}=\zeta_1$ and $\zeta_2^{\prime 5}=\zeta_2$. 
The monodromy transformation with respect to these coordinates is 
\begin{equation}
\label{montrsf}
\m:~(\zeta'_1,\zeta'_2)\rt(\omega\zeta'_1,\omega^2\zeta'_2).
\end{equation}
Monodromy of the solutions around the orbifold point is given by the matrix
representation of $\m$ in the basis of solutions obtained above.  

The monodromy matrix is obtained by expressing the transformation of the
periods under \eq{montrsf} in terms of the original periods. We first note
that, using the expressions derived above, 
under the monodromy transformation $\m$, 
$\Pi_3\rt\Pi_3-\frac{1}{5}\Pi_1$. That the 4-cycle under
monodromy goes over to a sum of the 2- and the 4-cycles has occurred in other
similar examples \cite{dg,mohri}. We fix the expression of $\Pi_4$, whose
analytic continuation has not been performed in the previous subsection, by
demanding that this 4-cycle volume under monodromy goes over to a combination
of $\Pi_4$ and $\Pi_2$ only.  This seems to be a
general feature for all $\Z_N$ orbifolds. This fixes $\Pi_4$. 
We rewrite the periods at the orbifold point in a convenient form as
\begin{equation}
\label{perorb}
\begin{split}
\varpi_0&=1,\\
\Pi_1 &=
\sum_{(m,n)'} 
(\omega^n - \omega^{4m+4n} - \omega^{2m} + \omega^{m+3n})\Pi(m,n),\\
\Pi_2 &=
\sum_{(m,n)'}
(\omega^{4m+4n} - 2\omega^{3m+2n} + \omega^{2m})\Pi(m,n),\\
\Pi_3 &=
-\frac{1}{5}\sum_{(m,n)'} 
(\omega^{4m+4n} - \omega^{m+3n})\Pi(m,n)+m_2,\\
\Pi_4 &=
-\frac{1}{5}\sum_{(m,n)'} 
( \omega^{3m+2n} -\omega^{2m}) \Pi(m,n)+m_1,
\end{split}
\end{equation} 
where we have defined 
\begin{equation}
\Pi(m,n) =-\frac{1}{(2\pi i)^3}
\frac{\Gm{\frac{m+2n}{5}}^2\Gm{\frac{3m+n}{5}}}{\Gm{1+m}\Gm{1+n}}
e^{\frac{i\pi}{5}(n-m)}\zeta_1^{\prime m}\zeta_2^{\prime n}.
\end{equation}

The monodromy matrix in the basis of periods
$(\varpi_0,\Pi_1 , \Pi_2 , \Pi_4 , \Pi_3)$ is obtained from these expressions
by applying $\mathfrak{m}$ on them and is given by           
\begin{equation}
{\frak m}=\begin{pmatrix}
1&0&0&0&0\\
5(m_1-2m_2)&-1&1&-5&10\\
5(m_2-3m_1)&1&-2&15&-5\\
0&0&-1/5&1&0\\
0&-1/5&0&0&1 
\end{pmatrix}. 
\label{Mo}\end{equation}
However, for the identification of fractional branes with
the wrapped branes, we
need to find out the branes wrapped on the cycles in the {\sc lcsl}, forming an
orbit of the monodromy matrix. Thus, we need the monodromy matrix in the basis
of periods in the {\sc lcsl}.
In the {\sc lcsl} basis $\varpi= (\varpi_0,\varpi_1, \varpi_2, \Pi_4, \Pi_3)$ 
of periods, the monodromy matrix
becomes\footnote{The fractions appearing in the monodromy matrix are due to 
the choice of the normalisation $1/5$ in the solutions quadratic in 
logarithms. These were chosen as unity in \cite{dg,mohri}.}, 
\begin{equation}
\M(\vpil{}{}) =\begin{pmatrix}
1&0&0&0&0\\
5(m_2-2m_1)&-2&1&10&-5\\
5(m_1-m_2)&1&-1&-5&5\\
0&-2/5&-1/5&1&0\\
0&-1/5&-3/5&0&1
\end{pmatrix}.
\label{Mvpil}\end{equation}
Due to the $\Z_5$ symmetry at the orbifold point, the monodromy matrix
satisfies ${\cal M}^5=\id{5}$. 
The canonical symplectic basis of periods in the {\sc lcsl}
admits a pairing through the symplectic intersection matrix 
\begin{equation}
{\cal I} = \begin{pmatrix}
0&0&0&0&0\\
0&0&0&1&0\\
0&0&0& 0&1\\
0&-1&0&0&0\\
0&0&-1&0&0
\end{pmatrix}. \label{intersection}
\end{equation}
The intersection pairing between the combinations of the periods in the {\sc lcsl},
forming an orbit of $\cal M$ yields the adjacency matrix of the quiver 
diagram of $\Z_5$, which we shall now discuss.
\subsection{Charges and brane configurations}
Having obtained the expressions of the periods \eq{perorb} at the orbifold
point, the monodromy $\cal M$
and their connection to the periods and monodromy in the {\sc lcsl}, we can now 
compare the D-brane configurations in these two regimes.
As mentioned before, at the orbifold point the basic objects are the 
fractional branes which are to be identified as bound states of wrapped 
branes in the large volume region. Due to the quantum $\Z_5$ 
symmetry at the orbifold point, the fractional branes
form an orbit of the monodromy. Thus, on identifying a single
fractional brane a set of five configurations can
be generated by successive right-action of the monodromy matrix on the charges 
of original wrapped brane. 

Let us start from a  D4-brane wrapped on the four-cycles, $\varpi_3$ and 
$\varpi_4$, such that the charge in the basis $\varpi$ of periods 
is ${\cal Q}_1 = (0,0,0,-1,2)$. This state has mass $2m_1-m_2$. 
Under the action of $\M(\varpi)$, this gives rise to the charges of 
four other states as ${\cal Q}_{i+1} = {\cal Q}_1 \M^i$, $i=1,2,3,4$.
The five states thus obtained are bound states of one wrapped D0-brane and two
D2- and two D4-branes, as can be identified from  $\varpi$. The states are 
tabulated in Table~\ref{frbrcharge}.
\begin{table}[h]
\begin{center}
\begin{tabular}{ccrrrr}
{\sc charge}&D0&$\text{D2}_1$&$\text{D2}_2$&$\text{D4}_1$&$\text{D4}_2$\\
&&&&&\\
${\cal Q}_1$ & 0&0&0&$-1$&$2$\\
${\cal Q}_2$ & 0&0&$-1$&$-1$&$2$\\
${\cal Q}_3$ & $5(m_2-m_1)$&$-1$&0&$4$&$-3$\\
${\cal Q}_4$ & $5m_1$&$1$&0&$-6$&$2$\\
${\cal Q}_5$ & $5(m_2-m_1)$&0&$1$&$4$&$-3$  
\end{tabular}
\end{center}
\caption{\small\sl Central charges corresponding to the bound states of branes}
\label{frbrcharge}
\end{table}
Each of these states has a mass $2m_1-m_2$, and since there are five states in
the orbit, by virtue of ${\cal M}^5=\id{5}$, 
we conclude that $2 m_1-m_2=1/5$, the mass of each fractional brane
in a unit in which the D0-brane has unit mass. A numerical evaluation of
$m_1$ and $m_2$ at one point of the principal component of the discriminant
locus \eq{PD} confirms this conclusion.

The pairing of the charges ${\cal Q}_i$, $i=1,{\ldots},5$,  using the
symplectic matrix \eq{intersection}, reproduces the adjacency 
matrix $\cal C$ of the quiver diagram of $\Z_5$, \ie
\begin{equation} 
C_{ij}={\cal Q}_i^{\mathrm{T}}{\cal I}{\cal Q}_j,
\end{equation}
where a superscript ${}^{\mathrm{T}}$ signifies matrix-transpose. The same has
also been obtained from the boundary states describing fractional branes
through the Witten index in \S\S\ref{witten}. 
Therefore, these five bound states of
D-branes given by the charges ${\cal Q}_i$ can be identified as the five
boundary states obtained there. In other words, this establishes the
connection between the irreducible representations of the group $\Z_5$ and the
combinations of the 0-, 2- and 4-cycle periods of the exceptional divisors 
of the crepant resolution of the orbifold $\C^3/\Z_5$, thus furnishing a
D-brane realisation of the McKay correspondence.

A further check on the choice of the initial charge ${\cal Q}_1$ is as
follows. The charges of the fractional branes ${\cal Q}_i$ have been 
conjectured \cite{dd,jay1,tom,mayr}
to be dual to the tautological line bundles ${\cal R}_i$ on the
resolution with respect to the pairing 
\begin{equation}
\int\ch{{\cal R}_i}{\cdot}\td{{\mathbf{X}}}{\cdot}\ch{{\cal Q}_j} = \delta{ij}, 
\end{equation}
where $\td{X}$ denotes the Todd class of $X$ and $\ch{{\cal E}}$ denotes 
the Chern character of a bundle $\cal E$. The charge vector
${\cal Q}_1$ chosen above may be checked to be dual to the trivial line
bundle with respect to this pairing.

A specification of brane configurations entails, in addition to the
dimensions of branes, the topological numbers associated to the gauge theory
supported on them. This information, referred to as the bundle data, is
obtained by comparing two expressions of the central charges \cite{mohri}
of the
supersymmetry algebra of the gauge theory, whose equality is necessitated by
anomaly considerations \cite{cyin,marco}.
One of the expressions for the central charges involves the periods and the
charges described above. For the case at hand, a generic bound state 
corresponds to 
\begin{dinglist}{235}
\item D4-branes wrapped on a surface in a homology class 
$(n^{[4]}_1,n^{[4]}_2)$ of the resolution
\item D2-branes corresponding to vector bundles with first Chern 
classes $(n^{[2]}_1, n^{[2]}_2)$  
\item D0-branes corresponding to an instanton number $n^{[0]}$. 
\end{dinglist}
On writing the charges in the basis $(t_{d1},t_{d2},t_1,t_2,1)$ of periods  
as $(n^{[4]}_1, n^{[4]}_2, n^{[2]}_1, n^{[2]}_2, n^{[0]})$,
the central charge of this configuration assumes the form
\begin{equation}
\sheaf{Q} = \sum_{i=1}^2n^{[4]}_i{\cdot}t_{di} 
+ \sum_{i=1}^2n^{[2]}_i{\cdot}t_i + n^{[0]}. 
\label{ccbrane}
\end{equation}
In large volume limit (or {\sc lcsl}), on the other hand, the central charge of a 
D4-brane wrapped on a surface $S$ with a vector bundle $\cal E$
can be obtained from considering cancellation of anomalies and is given by
\begin{equation}
\sheaf{Q} = \int_S e^{-\EuScript{K}}\ch{{\cal E}}
\sqrt{\frac{\hat{A}(T_S)}{\hat{A}(N_S)}},
\label{ccbundle}\end{equation}
where $\EuScript{K}$ is the K\"ahler form and $\hat{A}(T_S)$ and $\hat{A}(N_S)$ 
are the respective A-roof genus of the Tangent bundle and 
the Normal bundle of $S$ embedded in the resolution. 
The bundle data for the fractional branes can be obtained by equating
these two expressions of central charges.

In order to compare the two expressions \eq{ccbrane} and \eq{ccbundle} for the
central charge, we expand \cite{dg,mohri,sch,kllw} 
the K\"ahler form $\EuScript{K}$ in terms of $t_i$ as
$\EuScript{K}=t_1{\cdot}\mathfrak{d}_1+t_2{\cdot}\mathfrak{d}_2$.
Here we use the symbol $\mathfrak{d}_i$ also to denote the
restriction of divisor classes $\mathfrak{d}_i$  to $S$. 
Moreover, since $S$ is holomorphically embedded in the 
resolution $\mathbf{X}$ which has vanishing first Chern class, we have 
\begin{equation}
\sqrt{\frac{\hat{A}(T_S)}{\hat{A}(N_S)}} = \chi(S)\omega_S, 
\end{equation}
where $\chi(S)$ is the Euler number of the surface $S$ and 
$\omega_S$ satisfies $\int_S \omega_S = 1$. 
The expression \eq{ccbundle} thus reduces to a quadratic  
in $t_i$ as follows
\begin{equation}\label{ctq}
\sheaf{Q} = \frac{1}{2}\sum_{i,j=1}^2\int_S (\frak{d}_i{\cdot}
\frak{d}_j) t_i{\cdot}t_j 
-\sum_{i=1}^2\int_S\frak{d}_i{\cdot}\chun{{\cal E}}{\cdot}t_i 
-\frac{1}{24}\chi(S) 
+ \int_S\chdu{{\cal E}}.
\end{equation}
If $S$ belongs to a homology class $(n^{[4]}_1, n^{[4]}_2)$, then we have,
from the computation of the intersection numbers in \S\S\ref{yukaba}, the 
following expressions:
\begin{equation}
\begin{split}
\int_S \frak{d}_1^2 &= \frac{1}{5} (2n^{[4]}_1+n^{[4]}_2),\\
\int_S \frak{d}_2^2 &= \frac{3}{5} (n^{[4]}_1+3n^{[4]}_2),\\
\int_S (\frak{d}_2^2 &- 3\frak{d}_1{\cdot}\frak{d}_2) = 0.
\end{split}
\end{equation}
The first term in \eq{ctq}
can be written as a sum of $t_{d1}$ and $t_{d2}$ using the expressions
\eq{tds} and the intersection numbers. 
To simplify the second term, we write $\chun{{\cal E}} = a_1{\cdot}\frak{d}_1 
+ a_2{\cdot}\frak{d}_2$.  Substituting this, the second term of \eq{ctq}
becomes 
\begin{equation} 
\sum_{i,j,k}c_{ijk}a_jn^{[4]}_k{\cdot}t_i = \sum_in^{[2]}_i{\cdot}t_i.
\end{equation} 
The last term is, $n^{[0]} = \int_S\chdu{{\cal E}}$.
Combining these simplified expressions, we obtain, from \eq{ccbrane}, the
expression \eq{ccbrane} for the central charge $\sheaf{Q}$. 
This way of identifying the central charges \eq{ccbrane} and \eq{ccbundle}
leads to the 
explicit specification of the rank and Chern characters of the bundles
supported on the exceptional divisors on which the BPS branes are wrapped.
\section{Summary}\label{concl}
To summarise, in this article we have obtained the geometric realisations of
the boundary states corresponding to fractional D0-branes in ${\cal N}=2$ CFT
on the orbifold $\C^3/\Z_5$. The orbifold admits a crepant resolution.
The Newton polygon underlying this variety is not reflexive. 
The exceptional divisor of the resolution consists in $\cp{2}$ and
$\mathbb{F}_2$. By studying the periods of cycles of the resolution
with the help of 
the GKZ equations of its local mirror in the {\sc lcsl}, we obtained the 
B-type branes wrapped on the holomorphic cycles of the exceptional divisors. 
Analytic continuation of the
periods to the orbifold point provides the bound states of D0-, D2- and
D4-branes that form an orbit of the monodromy matrix at that point.
Identifying these five states as the five fractional branes by matching the
pairing matrix between these and the open-string Witten indices derived from
the CFT yields the large-volume brane configurations that correspond to the
fractional brane boundary states. Taking into account the RR-charges coupling
to the branes, this furnishes further evidence to identification of branes
as vector bundles on the cycles of the resolution.  

An understanding of D-branes wrapped on non-trivial cycles clarifies the 
behaviour of gauge theories with low supersymmetry in different phases 
and extends our understanding of D-branes. The identification of branes with
bundles on the cycles is important for obtaining a precise mathematical
definition of these objects. This precision also entails studies on
the stability of the bounds states, and hence of the associated bundles. This
generalises the concept of $\mu$-stability of vector bundles to
$\Pi$-stability. We have not considered the questions of stability of the bound
states in this article.  Such a study requires  explicit expressions of 
the periods near 
the singular loci. While possible in principle, this is technically
cumbersome, as we have seen already in \S\ref{orbsec}. We postpone it to a
future study.

It will be interesting to generalise these considerations to $G=\Z_N$ for
arbitrary $N$ and for non-abelian cases. Some 
mathematical results for these cases
are already available \cite{ito,reid}. This will also bring out a 
relationship between the generalised hypergeometric system and the
tautological bundles on the resolutions of the orbifolds, in view of 
more recent considerations that describes the fractional branes as duals of 
tautological bundles \cite{dd,jay1,tom,mayr}. 
\section*{Acknowledgement}
SM thanks Ansar Fayyazuddin and Alberto Guijosa for useful
discussions and comments. KR thanks M~Blau, E~Gava,
K~S~Narain and G~Thompson for helpful discussions and suggestions related to 
this work at various points during the work. 
We are grateful to Avijit Mukherjee for 
illuminating discourses throughout this period.
The research of SM is supported by a grant from the Swedish Research Council 
(NFR).

\end{document}